# Applications of statistical physics distributions to several types of income

**Elvis Oltean, Fedor V. Kusmartsev**

e-mail: elvis.oltean@alumni.lboro.ac.uk

**Abstract:** This paper explores several types of income which have not been explored so far by authors who tackled income and wealth distribution using Statistical Physics. The main types of income we plan to analyze are income before redistribution (or gross income), income of retired people (or pensions), and income of active people (mostly wages). The distributions used to analyze income distributions are Fermi-Dirac distribution and polynomial distribution (as this is present in describing the behavior of dynamical systems in certain aspects). The data we utilize for our analysis are from France and the UK. We find that both distributions are robust in describing these varieties of income. The main finding we consider to be the applicability of these distributions to pensions, which are not regulated entirely by market mechanisms.

## 1. Introduction

Papers that tackled the field of income and wealth distribution used most often so far as a measure for income disposable or net income. This implies that from gross income there are subtracted taxes and there are added transfers from public budget. We plan to see the scope of applicability to several types of income such as gross income (or income before redistribution), income of active people, wages and income of inactive people/pensions. However, these types of income that were measured are not fully equivalent in the cases of wages and income of active people. The most interesting and intriguing part about these data is that some of them depend partially or entirely on state and, therefore, market mechanisms play only a partial role. The only countries which provided this type of data regarding income apart from disposable income (divided in deciles) are France and the UK. We plan to use for the analysis of these types of income Fermi-Dirac and polynomial distributions.

## 2. Short Literature Review and Theoretical Framework

So far, previous research using these distributions applied to income and wealth distribution showed clearly that they are robust in their assessment. Fermi-Dirac distribution [1] was applied to income distribution for a group of countries mostly of them developed. The values for coefficient of determination were quite high and the temporal evolution of the coefficients obtained from fitting the data show some similarities with macroeconomic variables chosen to describe such systems. Also, in this paper a first analysis was made on other types of income apart from disposable income. The results showed clearly robustness of this distribution in the analysis of income distribution.

The polynomial distribution was first applied to the same data set as in the previous case and yielded similar results regarding the robustness [2]. The origin of this type of distribution comes from dynamic systems, where polynomials appear to model their behavior in certain aspects.

## 3. Methodology

Income can be measured using two types of calculation. Mean income is the sum of all individual incomes divided by number of inhabitants in a decile/population [3]. Upper limit on income is the highest individual or housedhold income from a decile of population. Deciles are segments of population divided equally in ten parts, where individuals or households are ranked increasingly according to their income. Thus, a decile contains 10% of actual population, the first one contain people with lowest income while the tenth decile contains people with highest income. Upper limit on income is a term coined by National Institute of Statistics from Finland [4]. In the case of the UK, the income is expressed in mean income. Instead, the data from France is expressed using upper limit on income, hence the upper income tier contained in the tenth decile is not considered. The data from France is expressed for individuals, while the ones for the UK are expressed for households.

In order to calculate the probability for a certain part of population to have an income, we use the cumulative probability density function. According to this type of probability, we calculate the share of population having an income above a certain threshold. Thus, the probability to have an income higher than zero is 100% (since everyone is assumed to have a certain income). Furthermore, in the case of the first decile the probability that people have a higher income is 90% and so forth. Let us assume $x_1, x_2, \ldots x_{10}$ be such that $x_1$ is the mean income for the first decile, and so forth up to $x_{10}$ which is the mean income for the tenth decile. The set which contains the plots representing the probability is S={ (0,100%), ($x_1$, 90%), ($x_2$, 80%), ($x_3$, 70%), ($x_4$, 60%), ($x_5$, 50%), ($x_6$, 40%), ($x_7$, 30%), ($x_8$, 20%), ($x_9$, 10%), ($x_{10}$, 0%)}. In order to fit the data, we used Fermi-Dirac distribution and polynomial distribution.

Fermi-Dirac distribution has its most general form as follows

$$\pi(\epsilon_i) = \frac{1}{exp\left(\frac{\epsilon_i - \mu}{T}\right) + 1} \quad (1)$$

where n represents the distribution of identical fermions over energy consisting of single-particle states. The parameters to be used for analysis are degeneracy (c), temperature (T), and chemical potential (μ). Therefore, the average number of fermions with a certain energy is calculated by multiplying *n* with degeneracy $g_i$ [5] such that

$$N(\epsilon_i) = \frac{g_i}{exp\left(\frac{\epsilon_i - \mu}{T}\right) + 1} \quad (2)$$

We represent graphically the results from Fermi-Dirac distribution using logarithmic values (log-log scale).

Polynomial distribution uses a slightly different approach by utilizing normal values in displaying the results. Unlike Fermi-Dirac distribution, the polynomial distribution can have variable number of coefficients ($P_1$, $P_2$, and/or $P_3$ as the case may be) according to the choice regarding the degree of the polynomial subject to the goodness of fit to the data. We chose the degree of the polynomials such that the values for the coefficient of determination (resulted from fitting the data) to be above 90%.

The categories of income taken into account are pensions, wages or active people income, and gross income (income before redistribution). In the case of pensions, the data from France are provided without making distinction between private and public pensions. In case of the UK, the data were provided separately for private and public pensions, and we added them up. It is noteworthy that taken separately they do not fit any distribution. Other types of income considered are income of active people or wages. However, we must show that these two types of income are not exactly the same as income of active people includes other types of income apart from wages such as income from self-employed activities. The third type of income considered is gross income or income before redistribution. This income represents the earnings of population before they pay taxes and receive benefits.

The results are obtained from fitting the data using Matlab toolbox for a confidence interval of 95%. The data we used are from the UK [6] and France [7], since these are the only countries which made these data available to the best of our knowledge.

## 4. Results

We show the results from fitting the annual data into the appendixes 1-10. In the appendixes 1-4, we exhibit the results from fitting the data using Fermi-Dirac distribution. The results from fitting the data using Fermi-Dirac distribution regarding pensions and income before redistribution for France were shown in [1]. The results from fitting the data using the polynomial distribution are displayed in the appendixes 5-10. In the following, we display graphically few typical examples of annual fitting to the data in the Figures 1-4. It is noteworthy that these examples were chosen such that the values for coefficient of determination for fitting the annual data to be the lowest for the each set of data for which results are displayed in each table.

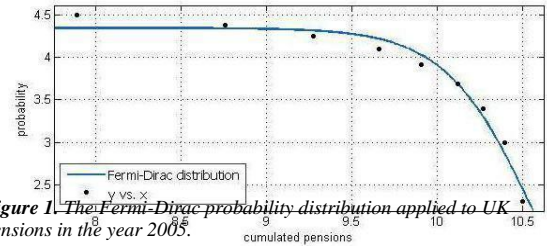

*Figure 1. The Fermi-Dirac probability distribution applied to UK pensions in the year 2005.*

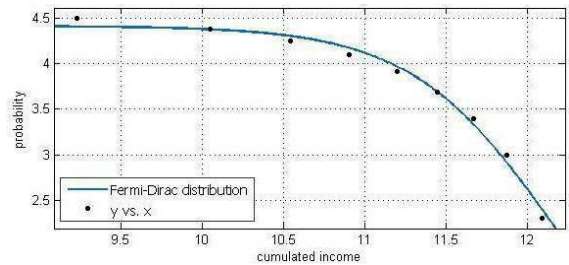

*Figure 2. The Fermi-Dirac probability distribution applied to active people's income from France in the year 2004.*

Generally, Fermi-Dirac distribution has better goodness of fit compared to polynomial distribution. A possible explanation is that Fermi-Dirac distribution has more coefficients than most of polynomial distributions we used.

Also, the higher the degree for polynomial distribution the higher are the annual values for coefficient of determination. Most of the data sets were described using first degree polynomials. The exception was about the UK wages, for which was possible to fit the data having annual values for coefficient of determination higher than 90% only by using second degree polynomials.

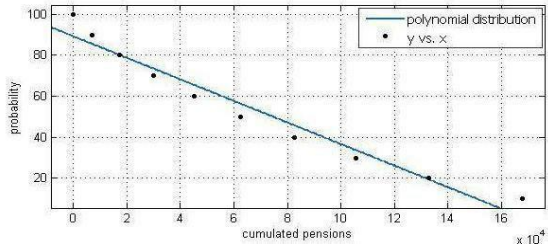

*Figure 3.* Polynomial probability distribution applied to pensions from France in the year 2003.

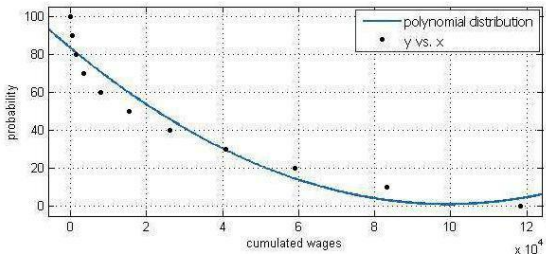

*Figure 4.* Polynomial distribution applied to UK wages in the year 1993.

These distributions are very robust. Thus, the lowest value for coefficient of determination in the case of Fermi-Dirac distribution is 97.3%. In the case of polynomial distribution, the lowest value regarding coefficient of determination is 89.7%. In the case of UK (which, unlike for France, contains the data regarding upper income segment of population), these distributions show that they are applicable to the entire range of income, namely to upper income segment of population which is thought traditionally to be described only by Pareto distribution.

It is very interesting that income of active people and gross income are described by market mechanisms, while pensions, even though obey the same distributions, they do not observe market mechanisms entirely. Thus, they are managed by public bodies and observe laws that follow some social principles which normally distort the market mechanisms.

## 5. Conclusions

Both Fermi-Dirac distribution and polynomial distribution describe with high degree of success different types of income. It is our opinion that other types of income can easily be described by these distributions.

The accuracy of polynomial distribution describing income distribution can be increased by using a higher degree polynomial, which implies an increased number of coefficients.

An important issue arises from the applicability of these distributions to other types of income depending entirely on market mechanisms and to others which, in addition, obey social principles carried out by taxes and benefits system. Further research should explain the share of influence of market mechanisms and social principle on the income distribution in general and on different categories of income in particular.

## Appendix

*Appendix 1.* Coefficients of the Fermi-Dirac distribution fitting income of active people in France.

| Year | T | C | μ | $R^2$ (%) |
|---|---|---|---|---|
| 2002 | 0.4483 | 4.411 | 12.18 | 98.98 |
| 2003 | 0.4441 | 4.41 | 12.18 | 98.96 |
| 2004 | 0.443 | 4.41 | 12.17 | 98.96 |
| 2005 | 0.4425 | 4.409 | 12.18 | 98.96 |
| 2006 | 0.4441 | 4.41 | 12.2 | 98.97 |
| 2007 | 0.4451 | 4.41 | 12.22 | 98.97 |
| 2008 | 0.4409 | 4.409 | 12.23 | 98.97 |
| 2009 | 0.4457 | 4.41 | 12.24 | 98.96 |

*Appendix 2.* Coefficients of the Fermi-Dirac distribution fitting gross income in the UK.

| Year | T | C | μ | $R^2$ (%) |
|---|---|---|---|---|
| 1977 | 0.4825 | 4.386 | 10.61 | 98.38 |
| 1978 | 0.4917 | 4.389 | 10.75 | 98.4 |
| 1979 | 0.4999 | 4.386 | 10.88 | 98.3 |
| 1980 | 0.4999 | 4.384 | 11.06 | 98.3 |
| 1981 | 0.5078 | 4.396 | 11.18 | 98.52 |
| 1982 | 0.5086 | 4.395 | 11.23 | 98.47 |
| 1983 | 0.5053 | 4.401 | 11.28 | 98.64 |
| 1984 | 0.5181 | 4.399 | 11.34 | 98.5 |
| 1985 | 0.5242 | 4.402 | 11.42 | 98.55 |
| 1986 | 0.5366 | 4.403 | 11.48 | 98.61 |
| 1987 | 0.5444 | 4.402 | 11.58 | 98.6 |
| 1988 | 0.5513 | 4.397 | 11.67 | 98.52 |
| 1989 | 0.5588 | 4.398 | 11.75 | 98.46 |
| 1990 | 0.5621 | 4.399 | 11.85 | 98.49 |
| 1991 | 0.563 | 4.4 | 11.91 | 98.54 |
| 1992 | 0.5558 | 4.404 | 11.94 | 98.55 |
| 1993 | 0.5601 | 4.411 | 11.95 | 98.73 |
| 1995 | 0.5591 | 4.41 | 12 | 98.63 |
| 1996 | 0.5498 | 4.41 | 12.03 | 98.73 |
| 1997 | 0.565 | 4.412 | 12.09 | 98.74 |

| Year | T | C | μ | $R^2$ (%) |
|------|------|------|------|------|
| 1998 | 0.5664 | 4.409 | 12.14 | 98.71 |
| 1999 | 0.5589 | 4.409 | 12.18 | 98.76 |
| 2000 | 0.5668 | 4.407 | 12.22 | 98.67 |
| 2001 | 0.5523 | 4.405 | 12.28 | 98.68 |
| 2002 | 0.5654 | 4.409 | 12.35 | 98.74 |
| 2003 | 0.5637 | 4.415 | 12.38 | 98.78 |
| 2004 | 0.5473 | 4.406 | 12.4 | 98.74 |
| 2005 | 0.5451 | 4.409 | 12.46 | 98.78 |
| 2006 | 0.5574 | 4.413 | 12.48 | 98.79 |
| 2007 | 0.5519 | 4.412 | 12.53 | 98.83 |
| 2008 | 0.5488 | 4.407 | 12.56 | 98.72 |
| 2009 | 0.5535 | 4.411 | 12.57 | 98.84 |
| 2010 | 0.5394 | 4.412 | 12.58 | 98.85 |
| 2011 | 0.5436 | 4.416 | 12.61 | 98.86 |
| 2012 | 0.5353 | 4.416 | 12.62 | 98.91 |

***Appendix 3.*** *Coefficients of the Fermi-Dirac distribution fitting pensions in the UK.*

| Year | T | C | μ | $R^2$ (%) |
|------|------|------|------|------|
| 1977 | 0.1714 | 4.363 | 8.283 | 97.69 |
| 1978 | 0.1861 | 4.378 | 8.448 | 98.09 |
| 1979 | 0.1553 | 4.361 | 8.578 | 98.15 |
| 1980 | 0.1631 | 4.372 | 8.736 | 98.37 |
| 1981 | 0.1912 | 4.363 | 8.94 | 97.89 |
| 1982 | 0.1886 | 4.359 | 8.994 | 98.07 |
| 1983 | 0.2375 | 4.365 | 9.163 | 97.67 |
| 1984 | 0.2159 | 4.366 | 9.206 | 98.35 |
| 1985 | 0.2398 | 4.376 | 9.331 | 98.56 |
| 1986 | 0.2346 | 4.368 | 9.396 | 98.15 |
| 1987 | 0.221 | 4.354 | 9.459 | 97.71 |
| 1988 | 0.2221 | 4.355 | 9.55 | 97.92 |
| 1989 | 0.2196 | 4.362 | 9.61 | 98.37 |
| 1990 | 0.2451 | 4.367 | 9.712 | 97.87 |
| 1991 | 0.2516 | 4.363 | 9.803 | 98.25 |
| 1992 | 0.2641 | 4.359 | 9.864 | 98.1 |
| 1993 | 0.281 | 4.356 | 9.939 | 97.8 |
| 1995 | 0.2933 | 4.374 | 10.02 | 98.8 |
| 1996 | 0.2862 | 4.362 | 10.07 | 97.77 |
| 1997 | 0.2967 | 4.36 | 10.09 | 97.85 |
| 1998 | 0.2997 | 4.36 | 10.17 | 97.99 |
| 1999 | 0.2755 | 4.352 | 10.22 | 97.85 |
| 2000 | 0.3013 | 4.362 | 10.29 | 97.79 |
| 2001 | 0.2843 | 4.36 | 10.33 | 0.98 |
| 2002 | 0.2604 | 4.342 | 10.4 | 97.12 |
| 2003 | 0.2827 | 4.35 | 10.48 | 97.78 |
| 2004 | 0.2868 | 4.355 | 10.55 | 97.79 |
| 2005 | 0.2606 | 4.343 | 10.57 | 97.3 |
| 2006 | 0.2955 | 4.363 | 10.61 | 98.13 |
| 2007 | 0.2851 | 4.353 | 10.65 | 97.82 |
| 2008 | 0.2847 | 4.355 | 10.68 | 97.99 |
| 2009 | 0.2942 | 4.349 | 10.74 | 97.55 |
| 2010 | 0.3127 | 4.35 | 10.83 | 97.31 |
| 2011 | 0.3288 | 4.355 | 10.9 | 97.5 |
| 2012 | 0.3352 | 4.353 | 10.92 | 97.52 |

***Appendix 4.*** *Coefficients of the Fermi-Dirac distribution fitting wages in the UK.*

| Year | T | C | μ | $R^2$ (%) |
|------|------|------|------|------|
| 1977 | 0.5801 | 4.361 | 10.34 | 97.98 |
| 1978 | 0.5974 | 4.364 | 10.53 | 98.01 |
| 1979 | 0.6171 | 4.355 | 10.6 | 97.79 |
| 1980 | 0.612 | 4.352 | 10.78 | 97.72 |
| 1981 | 0.6377 | 4.366 | 10.86 | 98.01 |
| 1982 | 0.6532 | 4.361 | 10.9 | 97.84 |
| 1983 | 0.6584 | 4.37 | 10.9 | 98.18 |
| 1984 | 0.6931 | 4.36 | 10.98 | 97.78 |
| 1985 | 0.6955 | 4.358 | 11.05 | 97.74 |
| 1986 | 0.7185 | 4.363 | 11.1 | 97.92 |
| 1987 | 0.7323 | 4.356 | 11.2 | 97.87 |
| 1988 | 0.7345 | 4.357 | 11.31 | 97.81 |
| 1989 | 0.7321 | 4.352 | 11.38 | 97.59 |
| 1990 | 0.7303 | 4.356 | 11.46 | 97.71 |
| 1991 | 0.7437 | 4.362 | 11.5 | 97.84 |
| 1992 | 0.7506 | 4.361 | 11.5 | 97.79 |
| 1993 | 0.7924 | 4.374 | 11.51 | 98.07 |
| 1995 | 0.7701 | 4.374 | 11.57 | 97.81 |
| 1996 | 0.7593 | 4.378 | 11.59 | 98.08 |
| 1997 | 0.7805 | 4.38 | 11.67 | 98.16 |
| 1998 | 0.7725 | 4.381 | 11.74 | 98.13 |
| 1999 | 0.7607 | 4.386 | 11.77 | 98.27 |
| 2000 | 0.7625 | 4.381 | 11.83 | 98.16 |
| 2001 | 0.7324 | 4.376 | 11.9 | 98.04 |
| 2002 | 0.7689 | 4.386 | 11.98 | 98.23 |
| 2003 | 0.7503 | 4.391 | 11.99 | 98.2 |
| 2004 | 0.7292 | 4.384 | 12 | 98.29 |
| 2005 | 0.7322 | 4.392 | 12.07 | 98.5 |
| 2006 | 0.7392 | 4.388 | 12.09 | 98.2 |
| 2007 | 0.7263 | 4.393 | 12.14 | 98.32 |
| 2008 | 0.7312 | 4.39 | 12.17 | 98.2 |
| 2009 | 0.7341 | 4.396 | 12.16 | 98.48 |
| 2010 | 0.7287 | 4.402 | 12.14 | 98.6 |
| 2011 | 0.7465 | 4.413 | 12.17 | 98.59 |
| 2012 | 0.7441 | 4.419 | 12.16 | 98.79 |

***Appendix 5.*** *Coefficients of the polynomials fitting income of active people in France.*

| Year | P1 | P2 | $R^2$ (%) |
|------|------|------|------|
| 2002 | -0.0004961 | 91.59 | 97.18 |
| 2003 | -0.0004976 | 91.76 | 97.3 |
| 2004 | -0.0005027 | 91.8 | 97.33 |
| 2005 | -0.0004954 | 91.77 | 97.33 |
| 2006 | -0.0004891 | 91.74 | 97.29 |
| 2007 | -0.0004774 | 91.69 | 97.26 |
| 2008 | -0.0004737 | 91.86 | 97.38 |
| 2009 | -0.0004677 | 91.65 | 97.25 |

***Appendix 6.*** *Coefficients of the polynomials fitting income before redistribution in France.*

| Year | P1 | P2 | $R^2$ (%) |
|------|------|------|------|
| 2002 | -0.00053 | 91.97 | 97.19 |
| 2003 | -0.0005338 | 92.15 | 97.29 |
| 2004 | -0.0005354 | 92.32 | 97.44 |
| 2005 | -0.0005296 | 92.19 | 97.34 |
| 2006 | -0.0005093 | 91.88 | 97.16 |
| 2007 | -0.0005079 | 92.01 | 97.25 |
| 2008 | -0.0005016 | 92.05 | 97.27 |
| 2009 | -0.0004986 | 92.03 | 97.28 |

***Appendix 7.*** *Coefficients of the polynomials fitting income of pensioners in France.*

| Year | P1 | P2 | $R^2$ (%) |
|---|---|---|---|
| 2003 | -0.0005256 | 89.16 | 95.78 |
| 2004 | -0.0005277 | 89.33 | 95.92 |
| 2005 | -0.0005197 | 89.27 | 95.87 |
| 2006 | -0.0005139 | 89.24 | 95.83 |
| 2007 | -0.0005057 | 89.22 | 95.84 |
| 2008 | -0.0004972 | 89.35 | 95.93 |
| 2009 | -0.0004928 | 89.15 | 95.82 |

***Appendix 8.*** *Coefficients of the polynomials fitting gross income in the UK.*

| Year | P1 | P2 | $R^2$ (%) |
|---|---|---|---|
| 1977 | -0.0021 | 85.79 | 94.45 |
| 1978 | -0.001833 | 85.57 | 94.27 |
| 1979 | -0.001604 | 85.07 | 93.99 |
| 1980 | -0.001328 | 84.79 | 93.71 |
| 1981 | -0.001175 | 85.03 | 93.48 |
| 1982 | -0.001116 | 84.98 | 93.44 |
| 1983 | -0.001058 | 85.16 | 93.23 |
| 1984 | -0.0009963 | 84.76 | 93.06 |
| 1985 | -0.0009065 | 84.32 | 92.43 |
| 1986 | -0.0008409 | 83.63 | 91.75 |
| 1987 | -0.0007615 | 83.14 | 91.34 |
| 1988 | -0.0006833 | 82.48 | 90.79 |
| 1989 | -0.0006413 | 82.56 | 91.11 |
| 1990 | -0.0005696 | 82.09 | 90.34 |
| 1991 | -0.0005349 | 82.13 | 90.43 |
| 1992 | -0.000527 | 82.7 | 90.91 |
| 1993 | -0.0005163 | 82.54 | 90.38 |
| 1995 | -0.0004942 | 82.7 | 90.65 |
| 1996 | -0.0004816 | 83.08 | 90.98 |
| 1997 | -0.0004505 | 82.44 | 90.44 |
| 1998 | -0.0004247 | 82.22 | 90.25 |
| 1999 | -0.0004047 | 82.24 | 90.1 |
| 2000 | -0.0003848 | 81.82 | 89.76 |
| 2001 | -0.0003668 | 82.44 | 90.5 |
| 2002 | -0.0003394 | 81.87 | 89.7 |
| 2003 | -0.0003348 | 82.6 | 90.61 |
| 2004 | -0.0003252 | 82.57 | 90.57 |
| 2005 | -0.0003088 | 82.86 | 90.79 |
| 2006 | -0.0002999 | 82.51 | 90.35 |
| 2007 | -0.0002842 | 82.58 | 90.38 |
| 2008 | -0.0002791 | 82.7 | 90.78 |
| 2009 | -0.000276 | 82.64 | 90.57 |
| 2010 | -0.0002716 | 83.01 | 90.68 |
| 2011 | -0.0002619 | 82.89 | 90.43 |
| 2012 | -0.0002646 | 83.51 | 91.14 |

***Appendix 9.*** *Coefficients of the polynomials fitting pensions in the UK.*

| Year | P1 | P2 | $R^2$ (%) |
|---|---|---|---|
| 1977 | -0.02404 | 107.2 | 96.93 |
| 1978 | -0.02064 | 107.5 | 97.34 |
| 1979 | -0.01792 | 108.7 | 95.77 |
| 1980 | -0.01523 | 108 | 95.91 |
| 1981 | -0.01241 | 105.1 | 98.09 |
| 1982 | -0.01169 | 104.4 | 98.03 |
| 1983 | -0.009715 | 100.3 | 99.56 |
| 1984 | -0.009348 | 102 | 99.06 |
| 1985 | -0.008257 | 100.7 | 99.39 |
| 1986 | -0.007774 | 101.1 | 99.49 |
| 1987 | -0.007186 | 100.9 | 99.5 |
| 1988 | -0.006549 | 100.8 | 99.52 |
| 1989 | -0.006227 | 101.5 | 99.3 |
| 1990 | -0.005682 | 100.3 | 99.68 |
| 1991 | -0.005139 | 99.34 | 99.8 |
| 1992 | -0.004765 | 97.67 | 99.83 |
| 1993 | -0.004392 | 96.02 | 99.73 |
| 1995 | -0.004116 | 96.62 | 99.78 |
| 1996 | -0.003866 | 96.02 | 99.71 |
| 1997 | -0.003782 | 95.65 | 99.69 |
| 1998 | -0.003487 | 95.09 | 99.61 |
| 1999 | -0.003322 | 96.16 | 99.75 |
| 2000 | -0.003124 | 95.3 | 99.67 |
| 2001 | -0.002992 | 96.41 | 99.8 |
| 2002 | -0.002779 | 96.63 | 99.72 |
| 2003 | -0.002558 | 95.69 | 99.72 |
| 2004 | 0.002435 | 96.15 | 99.77 |
| 2005 | -0.002368 | 97.16 | 99.79 |
| 2006 | -0.002251 | 95.48 | 99.7 |
| 2007 | -0.002161 | 95.62 | 99.72 |
| 2008 | -0.002103 | 95.72 | 99.7 |
| 2009 | -0.001981 | 94.85 | 99.6 |
| 2010 | -0.001826 | 94 | 99.44 |
| 2011 | -0.001692 | 93.2 | 99.26 |
| 2012 | -0.001645 | 92.53 | 98.98 |

***Appendix 10.*** *Coefficients of the polynomials fitting wages in the UK.*

| Year | P1 | P2 | P3 | $R^2$ (%) |
|---|---|---|---|---|
| 1977 | $7.233*10^{-8}$ | -0.004942 | 87.94 | 97 |
| 1978 | $5.013*10^{-8}$ | -0.004108 | 87.36 | 96.84 |
| 1979 | $4.474*10^{-8}$ | -0.003844 | 86.12 | 96.15 |
| 1980 | $3.066*10^{-8}$ | -0.00318 | 85.97 | 96.05 |
| 1981 | $2.773*10^{-8}$ | -0.003038 | 86.69 | 96.23 |
| 1982 | $2.607*10^{-8}$ | -0.00293 | 85.89 | 95.77 |
| 1983 | $2.628*10^{-8}$ | -0.002961 | 86.41 | 95.97 |
| 1984 | $2.314*10^{-8}$ | -0.002749 | 84.89 | 94.89 |
| 1985 | $1.994*10^{-8}$ | -0.002553 | 84.59 | 94.7 |
| 1986 | $1.815*10^{-8}$ | -0.002434 | 84.21 | 94.53 |
| 1987 | $1.524*10^{-8}$ | -0.002218 | 83.44 | 94.01 |
| 1988 | $1.229*10^{-8}$ | -0.001989 | 83.39 | 94.04 |
| 1989 | $1.057*10^{-8}$ | -0.001841 | 83.14 | 93.81 |
| 1990 | $8.876*10^{-9}$ | -0.001696 | 83.46 | 93.99 |
| 1991 | $8.219*10^{-9}$ | -0.001635 | 83.44 | 93.99 |
| 1992 | $8.217*10^{-9}$ | -0.001634 | 83.24 | 93.77 |
| 1993 | $8.323*10^{-9}$ | -0.001651 | 83.02 | 93.42 |
| 1995 | $7.375*10^{-9}$ | -0.001552 | 83.66 | 93.88 |
| 1996 | $7.072*10^{-9}$ | -0.001527 | 84.2 | 94.21 |
| 1997 | $6.199*10^{-9}$ | -0.001426 | 83.8 | 93.97 |
| 1998 | $5.224*10^{-9}$ | -0.001312 | 83.94 | 94.12 |
| 1999 | $4.683*10^{-9}$ | -0.001251 | 84.4 | 94.47 |
| 2000 | $4.329*10^{-9}$ | -0.001195 | 84.22 | 94.34 |
| 2001 | $3.703*10^{-9}$ | -0.001107 | 84.66 | 94.78 |
| 2002 | $3.159*10^{-9}$ | -0.001025 | 84.27 | 94.26 |
| 2003 | $3.101*10^{-9}$ | -0.001018 | 85.14 | 94.92 |
| 2004 | $2.921*10^{-9}$ | -0.0009893 | 85.09 | 95.13 |
| 2005 | $2.586*10^{-9}$ | -0.0009334 | 85.57 | 95.38 |
| 2006 | $2.511*10^{-9}$ | -0.0009175 | 85.23 | 94.98 |
| 2007 | $2.216*10^{-9}$ | -0.0008661 | 85.8 | 95.4 |
| 2008 | $2.12*10^{-9}$ | -0.000844 | 85.55 | 95.25 |
| 2009 | $2.174*10^{-9}$ | -0.0008579 | 85.81 | 95.36 |
| 2010 | $2.179*10^{-9}$ | -0.0008644 | 86.2 | 95.58 |
| 2011 | $2.129*10^{-9}$ | -0.000854 | 86.42 | 95.65 |
| 2012 | $2.131*10^{-9}$ | -0.0008583 | 86.75 | 95.76 |